\documentclass[final]{elsarticle} 

\usepackage{amssymb}
\usepackage{amsthm}
\usepackage{amsmath}
\usepackage{color}
\usepackage{mathtools}
\usepackage{graphicx}

\usepackage{caption}
\usepackage{subcaption}
\usepackage{subfig}
\usepackage{tikz}
\usetikzlibrary{arrows}
\usepackage{hhline}
\usepackage{multirow}
\usepackage{pbox}
\usepackage{makecell}
\usepackage{rotating}
\usepackage{tikz}

\newcommand{\calN}{\mathcal{N}}

\newcommand{\vecp}{\mathbf{p}}

\newcommand{\vecq}{\mathbf{q}}

\newcommand{\lm}{\textcolor{black}}

\newlength{\Oldarrayrulewidth}
\newcommand{\Cline}[2]{%
  \noalign{\global\setlength{\Oldarrayrulewidth}{\arrayrulewidth}}%
  \noalign{\global\setlength{\arrayrulewidth}{#1}}\cline{#2}%
  \noalign{\global\setlength{\arrayrulewidth}{\Oldarrayrulewidth}}}

\usepackage{nomencl}
\usepackage{etoolbox}
\renewcommand\nomgroup[1]{%
  \item[\bfseries
  \ifstrequal{#1}{R}{$\;\;\;\;\;\;$Roman:}{\ifstrequal{#1}{M}{$\;\;\;\;\;\;$Mathematical:}{\ifstrequal{#1}{A}{$\;\;\;\;\;\;$Abbreviations:}{\ifstrequal{#1}{G}{$\;\;\;\;\;\;$Greek:}{\ifstrequal{#1}{S}{$\;\;\;\;\;\;$Subscripts:}{\ifstrequal{#1}{T}{$\;\;\;\;\;\;$Superscripts:}{}}}}}}%
]}

\makenomenclature

\journal{Journal of Computational Physics}

\begin{document}

\begin{frontmatter}



\title{Stability analysis of thermo-acoustic nonlinear eigenproblems in annular combustors. \\Part II. Uncertainty quantification.}


\author[label1]{Luca Magri}
\author[label2]{Michael Bauerheim}
\author[label3]{Franck Nicoud}
\author[label4]{Matthew P. Juniper}

\address[label1]{Center for Turbulence Research, Stanford University, CA, United States of America}
\address[label2]{CERFACS, CFD Team, 31057 Toulouse, France}
\address[label3]{IMAG UMR-CNSR 5149, University of Montpellier, France}
\address[label4]{Cambridge University Engineering Department, Cambridge, United Kingdom}

\begin{abstract}
Monte Carlo and Active Subspace Identification methods are combined with first- and second-order adjoint sensitivities to perform (forward) uncertainty quantification analysis of the thermo-acoustic stability of two annular combustor configurations. This method is applied to evaluate the risk factor, i.e., the probability for the system to be unstable. It is shown that the adjoint approach reduces the number of nonlinear-eigenproblem calculations by up to $\sim\mathcal{O}(M)$, as many as the Monte Carlo samples.
\end{abstract}

\begin{keyword}
Thermo-acoustic stability \sep Uncertainty quantification \sep Annular combustors \sep Adjoint methods 
%
%
%
\end{keyword}

\end{frontmatter}
\renewcommand{\nomname}{Nomenclature}
\nomenclature[A]{UQ}{Uncertainty Quantification}
\nomenclature[A]{ASI}{Active Subspace Identification}
\nomenclature[A]{RF}{Risk Factor}
\nomenclature[A]{AD}{Adjoint}
\nomenclature[A]{FD}{Finite difference}
\nomenclature[A]{PDF}{Probability Density Function}
\nomenclature[A]{MC}{Standard Monte Carlo method (benchmark solution)}

\nomenclature[G]{$\omega$}{Complex eigenvalue, $\omega_r+\mathrm{i}\omega_i$}
\nomenclature[G]{$\omega_r$}{Angular frequency}
\nomenclature[G]{$\omega_i$}{Growth rate}
\nomenclature[G]{$\omega_i^{ASI}$}{Growth rate by surrogate models}
\nomenclature[M]{$\left\langle\cdot,\cdot\right\rangle$}{Inner product}
\nomenclature[M]{$\mathcal{N}$}{Operator representing the nonlinear eigenvalue problem}
\nomenclature[R]{$\mathbf{p}$}{Vector of thermo-acoustic parameters}
\nomenclature[R]{$\vecq$}{State vector}
\nomenclature[R]{$M$}{Monte Carlo samples for uncertainty quantification of $\omega_i$}
\nomenclature[R]{$M^{cov}$}{Monte Carlo samples to generate the uncentred covariance matrix in ASI}
\nomenclature[R]{$M^{ASI}$}{Monte Carlo samples to develop the surrogate models with ASI by regression}
\nomenclature[R]{$N$}{Eigenvalue geometric degeneracy}
\nomenclature[R]{$\mathrm{i}$}{Imaginary unit, $\mathrm{i}^2+1=o$}
\nomenclature[R]{$N_{reg}$}{Number of regression coefficients}

\nomenclature[S]{$0$}{Unperturbed}
\nomenclature[S]{$1$}{First-order perturbation}
\nomenclature[S]{$2$}{Second-order perturbation}
\nomenclature[T]{$H$}{Hermitian}
\nomenclature[T]{$T$}{Transpose}
\nomenclature[T]{$*$}{Complex conjugate}
\nomenclature[T]{$\hat{}$}{Eigenfunction}
\nomenclature[T]{$+$}{Adjoint}

\printnomenclature[10mm]
\section{Introduction}\label{sec:intro}
Thermo-acoustic oscillations involve the interaction of heat release (e.g., from a flame) and sound. In rocket and aircraft engines, as well as power-generation turbines, heat release fluctuations can synchronize with the natural acoustic modes in the combustion chamber. This can cause large oscillations of the fluid quantities, such as the static pressure, that sometimes lead to catastrophic failure. It is one of the biggest and most persistent problems facing rocket \citep{Culick2006} and aircraft engine manufacturers \citep{Lieuwen2005}. 

\begin{figure}
  \begin{center}  
 \includegraphics[scale=0.5]{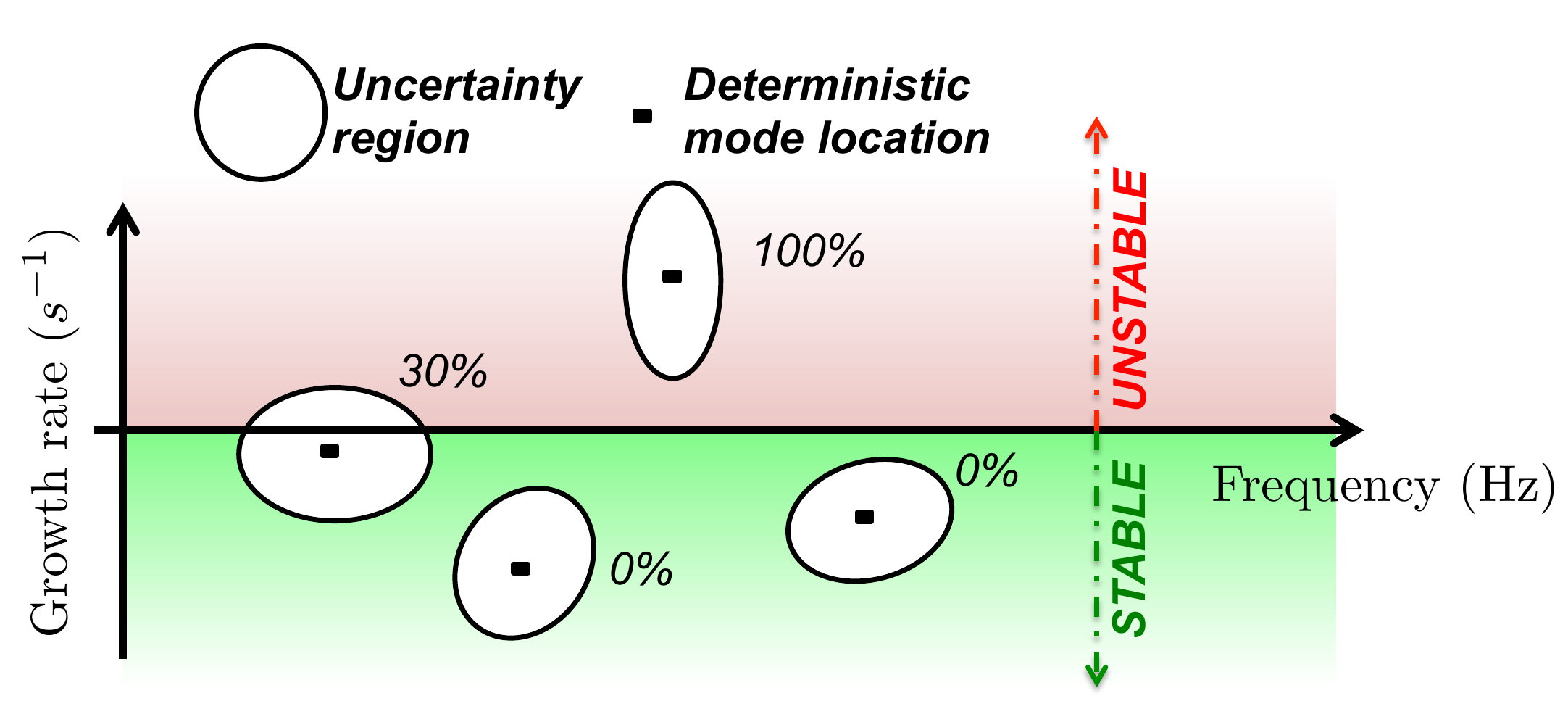}
  \end{center}
  \caption{Pictorial view of the deterministic locations (black symbols) in the complex plane of the first four thermo-acoustic eigenvalues in a typical combustor. When uncertainties are taken into account, the eigenvalues belong to an admissible region of the complex plane (circles) associated with a risk factor, which describes the percentage probability that the mode is unstable.}\label{fig:uqstab}
  \end{figure}

The output of any frequency-based stability tool is usually a map of the thermo-acoustic eigenvalues in 
the complex plane (black squares in Fig.~\ref{fig:uqstab}).  
Each thermo-acoustic mode must have negative growth rate for the combustor to be linearly stable. 
The design process is even more complex because of the uncertainty in the thermo-acoustic parameters $\mathbf{p}$ of the low-order thermo-acoustic model. For example, the speed of sound, the boundary impedances and the flame model are sensitive to partly unknown physical parameters such as the flow regime, manufacturing tolerances, fuel changes, or acoustic and heat losses. As a consequence, each mode actually belongs to an uncertain region of the complex plane (Fig.~\ref{fig:uqstab}). This uncertain region is measured by the risk factor~\cite{Bauerheim2016}, which corresponds to the probability that the mode is unstable. Although the probabilistic estimation (uncertainty quantification) of the thermo-acoustic stability is paramount for practitioners, there are only a few studies in the literature \cite{Hagen2004,Bauer2014ctr,Ndiaye2015,Bauerheim2016}. 

Uncertainty Quantification (UQ) of thermo-acoustic stability was performed in longitudinal academic configurations containing one turbulent flame in Ndiaye et al. \cite{Ndiaye2015}. Assuming that only the flame model was uncertain, i.e., the gain and the time delay, the risk factor of the system was obtained by combining a Helmholtz solver \citep{Nicoud2007} with a Monte-Carlo analysis. Each computation required a few tens of minutes making the generation of a 10,000 Monte-Carlo sample database CPU-demanding. 
To reduce the CPU cost of UQ analysis, network models \citep{Evesque2002,Dowling1995} can be used, especially for cases involving many uncertain parameters, such as multiple-flame configurations in annular combustors \cite{Hagen2004}. Low-order models are suitable for studying how the uncertainties in the input parameters propagate and affect the uncertainties in the eigenvalues ({\it forward} UQ as defined in Chantrasmi and Iaccarino \cite{Chantrasmi2012}) . This was performed by Bauerheim et al. \cite{Bauerheim2016} who applied a standard Monte Carlo analysis to a 19-burner annular configuration represented by a network-based model with 76 acoustic elements and subsequently reduced to a 4$\times$4 matrix through Annular Network Reduction \cite[]{Bauerheim2014a}. Assuming that only the amplitude and phase of the 19 flame responses were uncertain, they found that approximately 10,000 computations were necessary to assess the risk factor of the annular combustor.

In order to avoid expensive Monte Carlo methods and speed up the uncertainty evaluation, a UQ approach called Active Subspace Identification (ASI), as proposed by Constantine et al. \cite[]{Constantine2014} and Lukaczyk et al. \cite{Lukaczyk2014}, was tested in Bauerheim et al. \cite{Bauer2014ctr,Bauerheim2016}. The objective was to reduce the dimension of the parameter space  to just a few by analysing growth-rate gradients, $\partial \omega_i / \partial \vecp$. 
A set of ``active variables" were then calculated to describe the response surface of the growth rate, i.e., the function $\omega_i=\omega_i(\mathbf{p})$ by least-square methods. 
In the annular combustor investigated by \cite{Bauerheim2016}, only three/five active variables were sufficient to represent the 38-dimensional response surface with surrogate algebraic models obtained by regression. 
Using these surrogate models, they performed a Monte Carlo analysis at lower cost to calculate the risk factors given uncertainties in the input parameters. 
Evaluating the gradients $\partial \omega_i / \partial \vecp$ \lm{by finite difference} is a time-consuming task when the number of parameters, $\vecp$, and the Monte Carlo sampling are large. Consequently, being able to accurately estimate the gradients of the growth rate $\omega_i(\vecp)$ at low cost is necessary to achieve an efficient UQ analysis \cite{MagriPhD,Juniper2014ctr}. 

The aim of this paper is to reduce such a computational effort by combining first- and second-order adjoint-based eigenvalue sensitivities (Section \ref{sec:appannu}), detailed in Part I of this paper \cite{Magri2016jcp1},  with a standard Monte Carlo method (Section \ref{sec:monca}) and a Monte Carlo method integrated with Active Subspace Identification (Section \ref{sec:asm}) to predict the probabilities that two annular-combustor configurations are unstable. 

\section{Mathematical framework}\label{sec:appannu}
\begin{figure}[!htb]
  \begin{center} 
 \includegraphics[scale=0.45]{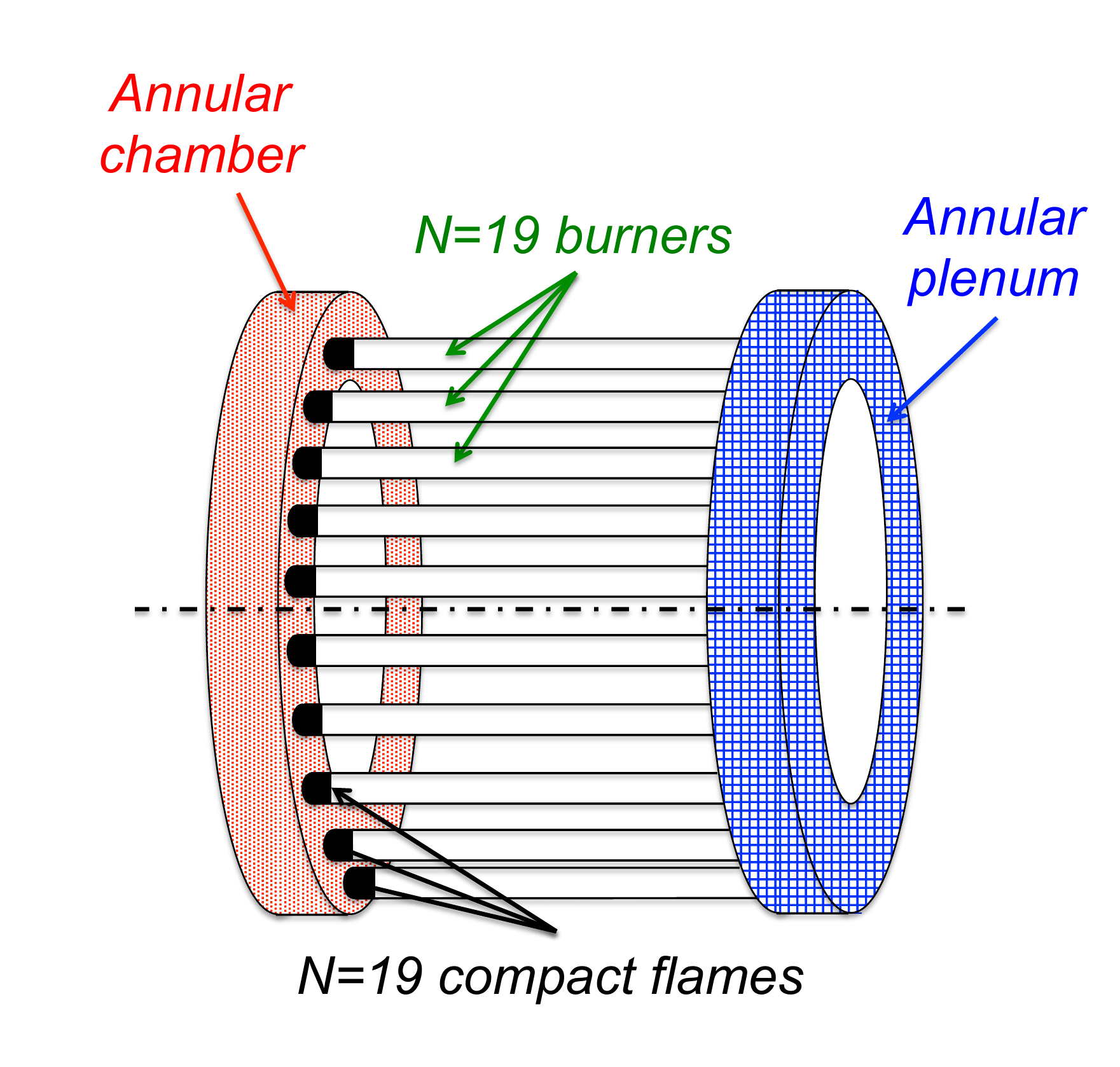}
  \end{center}
  \caption{Schematic of the annular combustor, which consists of a plenum and combustion chamber connected by longitudinal burners \citep{Bauerheim2014a,Bauerheim2016}. }\label{fig:network}
\end{figure}

We study the same annular combustor of Part I of this paper, whose model is detailed in \cite{Bauerheim2014a}. It consists of a combustion chamber connected by longitudinal burners fed by a common annular plenum (Fig. \ref{fig:network}). 

We briefly recall the theoretical framework that we need here. 
The stability is governed by a nonlinear eigenproblem\footnote{From now on, the subscript $0$ denotes unperturbed quantities, $1$ denotes first-order quantities and $2$ denotes second-order quantities.} 
\begin{align}\label{eq:eig1}
&  \calN\left\{\omega_0,\mathbf{p}_0\right\}\hat{\vecq}_0=0, 
\end{align}
where 
$\omega_0$ is the eigenvalue\footnote{Its real part, $\omega_r$, represents the angular frequency (rad/s) whereas the imaginary part, $\omega_i$, represents the growth rate (1/s).}, which appears under nonlinear terms of exponential, polynomial and rational type,  and $\hat{\vecq}_0$ is the eigenfunction containing the acoustic pressure and velocity at two locations in the plenum and combustion chamber (see \cite{Bauerheim2016} for details).  Here, the vector of thermo-acoustic paramaters contains only the flame parameters, $\mathbf{p}_0$ $=$ $(\{n_{0i},$ $\tau_{0i}\})$, where $i=1,2,\ldots,N$, because they are assumed to be the most uncertain factors \cite{Bauerheim2016}. The flame gains (or indices), $n_{0i}$, and the time delays, $\tau_{0i}$, relate the unsteady heat-release rate, $\hat{\mathcal{Q}}$, to the acoustic velocity, $\hat{u}_i$, at the burner's location as a Flame Transfer Function \cite{Crocco1956}
\begin{align}
\hat{\mathcal{Q}}=\frac{\bar{p}\gamma}{\gamma-1}S_in_{0i}\exp(\mathrm{i}\omega_0\tau_{0i})\hat{u}_i,
\end{align}
where $\gamma$ is the unburnt-gas heat capacity ratio, $\bar{p}$ is the static pressure and $S_i$ is the section of the $i$th burner. The combustor has 19 burners (Fig.~\ref{eq:eig1}), thus, 38 flame parameters. 
These parameters are reported in Tables 1 and 2 in Part I. 

We perturb a flame parameter, $\mathbf{p}=\mathbf{p}_0+\epsilon\mathbf{p}_1$, and calculate the perturbation operator numerically as $\delta_p\calN\{\omega_0,\epsilon\mathbf{p}_1\}=  \calN\{\omega_0,\mathbf{p}\} -\calN\{\omega_0,\mathbf{p}_0\}$, where  $\epsilon\ll 1$.
In Part I, it was shown that the first-order eigenvalue drift reads 
\begin{align}\label{2eq:eigpert5}
& \omega_1 = 
 \frac{- \left\langle\hat{\vecq}_0^+, \delta_p\calN\left\{\omega_0, \epsilon\mathbf{p}_1\right\}\hat{\vecq}_0\right\rangle}{\left\langle\hat{\vecq}_0^+, \frac{\partial\calN\left\{\omega, \mathbf{p}_0\right\}}{\partial\omega}\Big\lvert_{\omega_0}\hat{\vecq}_0\right\rangle}, 
\end{align}
where $\hat{\vecq}_0^{+}$ is the adjoint eigenfunction, which is solution of the adjoint eigenproblem $\calN\left\{\omega_0,\mathbf{p}_0\right\}^H\hat{\vecq}_0^{+}$$=$$0$, in which  $\langle\cdot,\cdot\rangle$ represents an inner product and $^H$ is the complex transpose. 
If the unperturbed eigenvalue $\omega_0$ is $N$-fold degenerate, and $\hat{\mathbf{e}}_{0,i}$ are the $N$ independent eigenfunctions associated with it,  we obtain an eigenproblem for the first-order eigenvalue drift, $\omega_1$, and eigendirection, $\alpha_j$, as follows 
\begin{align}\label{eq:degenerate}
& \left\langle\hat{\mathbf{e}}^+_{0,i},\frac{\partial\calN\left\{\omega, \mathbf{p}_0\right\}}{\partial\omega}\Big\lvert_{\omega_0}\hat{\mathbf{e}}_{0,j}\right\rangle\omega_1\alpha_j = -\left\langle\hat{\mathbf{e}}^+_{0,i}, \delta_p\calN\left\{\omega_0, \epsilon\mathbf{p}_1\right\}\hat{\mathbf{e}}_{0,j}\right\rangle\alpha_j,
\end{align}
for $i,j=1,2,...,N$. Einstein summation is used, therefore, the inner products in equation \eqref{eq:degenerate} are the components of an $N\times N$ matrix and $\alpha_j$ are the components of an $N\times 1$ vector.  
Among the $N$ eigenvalues drifts, $\omega_1$, outputted by \eqref{eq:degenerate}, we select the one with greatest growth rate because it causes the greatest change in the stability. In thermo-acoustics, degeneracy occurs in rotationally symmetric annular combustors in which azimuthal modes have 2-fold degeneracy \cite[see, e.g.,][]{Taylor2011}. 

The second-order eigenvalue drift reads 
\begin{align}\nonumber
& \omega_2 = 
-  2\frac{\left\langle\hat{\vecq}^+_0,\left(\frac{\partial\calN\left\{\omega, \mathbf{p}_0\right\}}{\partial\omega}\Big\lvert_{\omega_0}\omega_1\hat{\vecq}_1+ \delta_p\calN\left\{\omega_0, \epsilon\mathbf{p}_1\right\}\hat{\vecq}_1\right)\right\rangle}{\left\langle\hat{\vecq}^+_0,\frac{\partial\calN\left\{\omega,\mathbf{p}_0\right\}}{\partial\omega}\Big\lvert_{\omega_0}\hat{\vecq}_0\right\rangle} + \\
& -2\frac{\left\langle \hat{\vecq}^+_0,\left(\frac{1}{2}\frac{\partial^2\calN\left\{\omega,\mathbf{p}_0\right\}}{\partial\omega^2}\Big\lvert_{\omega_0}\omega_1^2  \right)\hat{\vecq}_0\right\rangle}{\left\langle\hat{\vecq}^+_0,\frac{\partial\calN\left\{\omega,\mathbf{p}_0\right\}}{\partial\omega}\Big\lvert_{\omega_0}\hat{\vecq}_0\right\rangle}.\label{2eq:eigpert8}
\end{align}
The calculation of the perturbed eigenfunction $\hat{\vecq}_1$, which is necessary only for the calculation of the second-order eigenvalue drift, is described in \cite{Magri2016jcp1}. 
\section{Uncertainty quantification via standard Monte Carlo method}\label{sec:monca}
Part I of this paper showed that using the adjoint method can drastically reduce the computational cost of deterministic sensitivity analysis. A similar approach can, thus, be used when the parameters are varied randomly. This section shows how the adjoint method can provide efficient Uncertainty Quantification (UQ)  strategies to predict thermo-acoustic stability from a probabilistic standpoint.
We study two of the three configurations of Part I\footnote{Case B is the rotationally symmetric version of Case C. They have a similar probabilistic behaviour (not shown) and Case B is not reported here for brevity.}, i.e., the weakly-coupled rotationally symmetric  Case A, and the strongly-coupled rotationally asymmetric Case C, which is relevant to industrial configurations \cite{Campa2011}.
A standard Monte Carlo method (MC) is integrated with the adjoint formulation for the  calculation of the Probability Density Function (PDF) and Risk Factor (RF), the latter of which is defined as the probability that the system is unstable, given a PDF for the input parameters \cite{Bauerheim2016}
\begin{align}\label{eq:riskfactor}
& \textrm{RF} = \int_{0}^{\infty}\textrm{PDF}(\omega_i)\textrm{d}\omega_i. 
 \end{align}
In practical applications, the uncertainties are typically greatest in the flame parameters \cite{Bauerheim2016}. Therefore we calculate how the thermo-acoustic growth rate, $\omega_i$, which governs the stability, is affected by uncertain flame parameters. Studying uncertainty quantification for other parameters is just as straightforward. 
We assume we know the maximum and minimum values of the uncertain flame parameters. Using the Principle of Maximum Entropy, we choose the uniform distribution for the input parameters because it is the least biased possible distribution given the available information \cite{Jaynes1957}. 
Note that~\cite{Ndiaye2015} have shown that the PDF shape has a minor effect on the risk factor in the case they considered.

By the standard Monte Carlo method used by \cite{Bauerheim2016},  
$M$ random values of $\vecp$, called the Monte Carlo sampling $\vecp^{MC}$, are selected with respect to their PDFs and the nonlinear eigenproblem \eqref{eq:eig1} is solved $M$ times to provide $M$ eigenvalues. This means that with this method, which is the reference solution, we have to solve for $M$ nonlinear eigenproblems. The Monte Carlo method always converges to the final PDF but suffers from slow convergence, being $\sim\mathcal{O}(1/\sqrt{M})$, which could be prohibitive in large systems such as the Helmholtz equation in complex geometries. This calls for the adjoint-based method. 
  
To avoid the computations of the $M$ samples, here $\sim\mathcal{O}(10^4)$, the Monte Carlo analysis is viewed as a random perturbation around the unperturbed state. Consequently, the adjoint sensitivities of Section~\ref{sec:appannu} are applied to obtain the eigenvalue drifts providing an efficient UQ strategy. Thus, the random sequence of parameters $\vecp^{MC}$ is used as a perturbation in  the first-order eigenvalue drift in equation \eqref{eq:degenerate}, for degenerate eigenproblems, or in equation \eqref{2eq:eigpert5} for non-degenerate cases. Then, the perturbed eigenvector is calculated by SVD and the second-order drift is calculated by equation \eqref{2eq:eigpert8} for each sequence of random parameters. 
These are only vector-matrix-vector multiplications or lower-rank linear systems. Importantly,  the adjoint method requires the computation of only one nonlinear eigenproblem \eqref{eq:eig1} and its adjoint, regardless of the Monte Carlo sampling $M$ or the number of perturbed parameters.

First, we evaluate how many Monte Carlo samples are needed for the risk factors to converge. 
Table \ref{tab:conv} shows the convergence of the risk factor for three Monte Carlo samplings of $10,000$, $20,000$ and $30,000$ imposing a standard deviation of $5\%$ on the flame indices and time delays. We choose the Monte Carlo sampling of $M=10,000$ for UQ as a compromise between accuracy and computational cost. 

\begin{table}[!htb]
\centering
\renewcommand{\arraystretch}{1.3} 
  {\small{\begin{tabular}{ c | c | c | c |  c   }
      & \makecell{Monte Carlo\\ samples} & RF via MC         & RF via 1st-order AD   & RF via 2nd-order AD   \\ \Cline{2pt}{1-5}
      \parbox[t]{2mm}{\multirow{3}{*}{\rotatebox[origin=c]{90}{Case A}}} & 10,000                               &     33.3\%                                                  &40.3\%         &       34.9\%            \\  
        & 20,000                               &     33.9\%                                                  & 41.3\%             & 34.2\%              \\  
       & 30,000                               &     33.3\%                                                  & 40.6\%                       &  33.7\%   \\ \Cline{1pt}{1-5}
      \parbox[t]{2mm}{\multirow{3}{*}{\rotatebox[origin=c]{90}{Case C}}}  &10,000                               &     40.9\%                                                  &34.7\%                    &    41.5\%    \\  
        &20,000                               &     40.8\%                                                  & 34.5\%                    &    41.3\%   \\  
        &30,000                               &     40.9\%                                                  & 34.6\%                     &    41.1\%   \\  
  \end{tabular}}}
\caption{Risk factors (RF) calculated by the Monte Carlo method as a function of the Monte Carlo samples. MC is the standard Monte Carlo method, which provides the benchmark solution, and AD stands for adjoint. The standard deviation of the flame indices and time delays is $5\%$.}\label{tab:conv} 
\end{table}

Secondly, we impose different standard deviations to the uniform distributions of the flame parameters and calculate the growth-rate PDFs and risk factors. 
The results are shown in Table \ref{tab:RF}. When the standard deviations are smaller than $2.5\%$, the first-order adjoint method provides accurate predictions, although it becomes less accurate for larger deviations. However, the second-order adjoint method provides accurate predictions of the risk factor up to standard deviations of $10\%$, matching satisfactorily the benchmark solution by MC. 

\begin{table}[!htb]
\centering
\renewcommand{\arraystretch}{1.3} 
  {\small{\begin{tabular}{ c | c | c |  c |  c }
       & \makecell{Standard\\ deviation} & RF via MC        & RF via 1st-order AD       & RF via 2nd-order AD \\ \Cline{2pt}{1-5}
       \parbox[t]{2mm}{\multirow{4}{*}{\rotatebox[origin=c]{90}{Case A}}} & 1\%                               &     15.4\%                                                  &14.7\%        &                 15.4\%   \\  
        & 2.5\%                            &     31.3\%                                                  & 33.2\%         &      31.2\%            \\  
        & 5\%                               &     33.25\%                                                & 40.3\%           &     33\%            \\  
        & 10\%                             &     34.5\%                                                  & 43.2\%             &    34.9\%          \\  \Cline{1pt}{1-5}
          \parbox[t]{2mm}{\multirow{4}{*}{\rotatebox[origin=c]{90}{Case C}}}&1\%            &     18.3\%           &17.5\%     &    18.3\%             \\  
        &2.5\%                            &     35.4\%                                                  & 31.1\%            &     35.3\%          \\  
        &5\%                               &     40.9\%                                                  & 34.7\%                &     40.4\%       \\  
        &10\%                             &     42.2\%                                                  &30.0 \%                &      41.5\%     \\ 
  \end{tabular}}}
\caption{Risk factors (RF) calculated by the Monte Carlo method as a function of the standard deviation of the flame index and time delay uniform distributions. MC is the standard Monte Carlo method and AD stands for adjoint. }\label{tab:RF} 
\end{table}

\begin{figure}[!htb]
        \centering
        \includegraphics[width=0.9\linewidth]{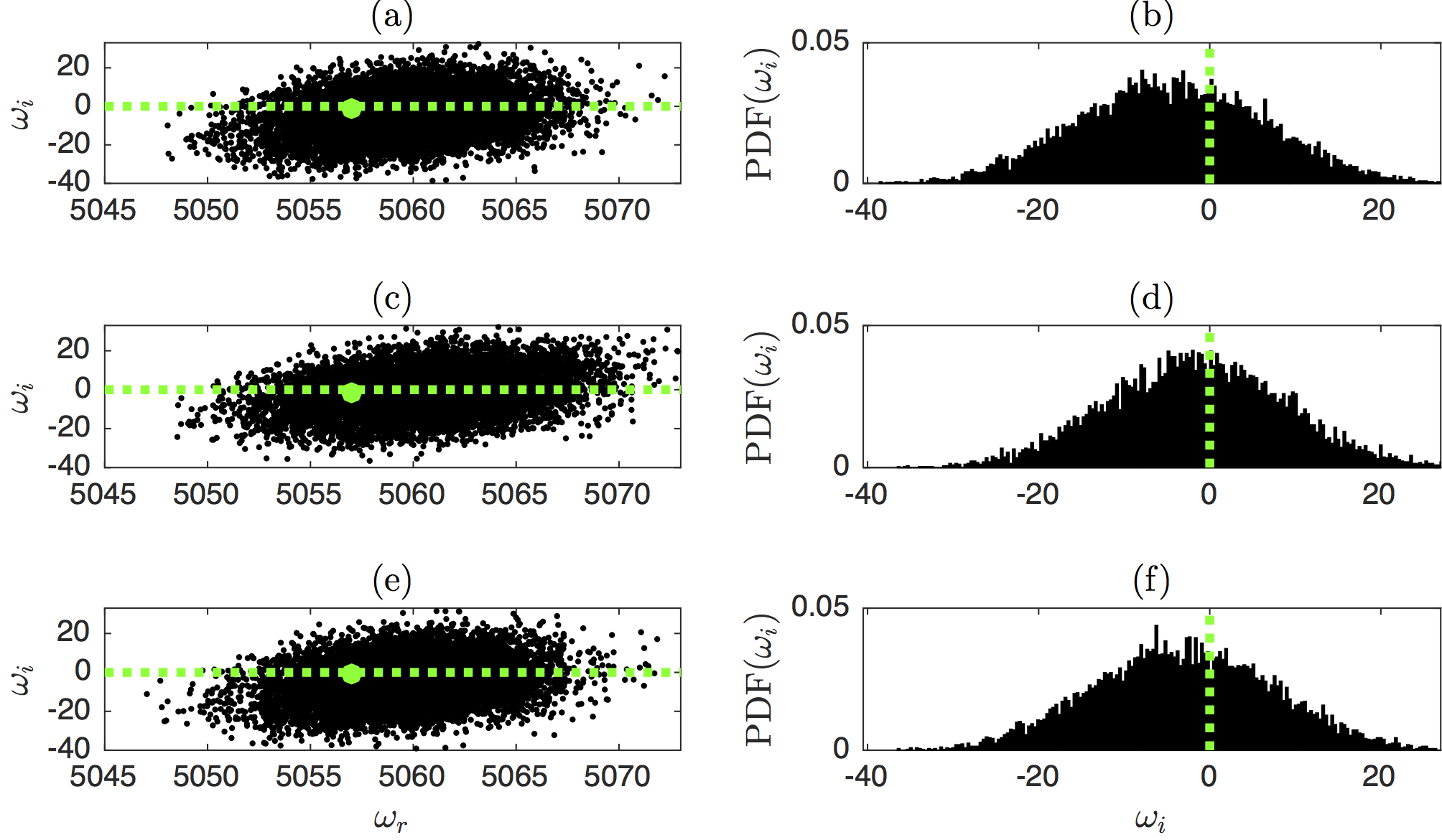}
  \caption{Weakly coupled Case A. Perturbed eigenvalues calculated by the standard Monte Carlo method (MC, first row), 1st-order adjoint method (second row) and  2nd-order adjoint method (last row) via a Monte Carlo sampling of 10,000. With the MC approach, 10,000 nonlinear eigenvalue problems are solved, whereas with the AD approaches only one eigenproblem and its adjoint are solved. Normalized histograms of the growth rates are shown in the right panels. The dotted lines divide the stable plane ($\omega_i<0$) from the unstable plane ($\omega_i>0$). The big dot is the unperturbed deterministic eigenvalue, $\omega_0=5.059\times10^3rad/s-\mathrm{i}1.392s^{-1}$. The standard deviation of the uniform distribution of the flame parameters is $10\%$ (last row of Table \ref{tab:RF} for Case A).}\label{fig:MCa}
\end{figure}

In Fig. \ref{fig:MCa} we depict the eigenvalues via Monte Carlo simulations obtained by MC (first row), first-order adjoint method (middle row) and second-order adjoint method (bottom row) for the weakly coupled Case A.  The  clouds (left panels) are obtained by imposing a uniform probability distribution between $\pm0.1n, \pm0.1\tau$, which represent the uncertainties of the flame parameters (last row of Table \ref{tab:RF} for Case A). The PDFs of the perturbed growth rates are depicted in the right panels. The PDF shape is satisfactorily predicted by the first-order adjoint method, however, to obtain accuracy on the risk factor the second-order adjoint formulation is necessary. For this case, the risk factor predicted by MC is 34.5\%, by first-order AD is 43.2\% and by second-order AD is $34.9\%$.  

\begin{figure}[!htb]
        \centering
        \includegraphics[width=0.9\linewidth]{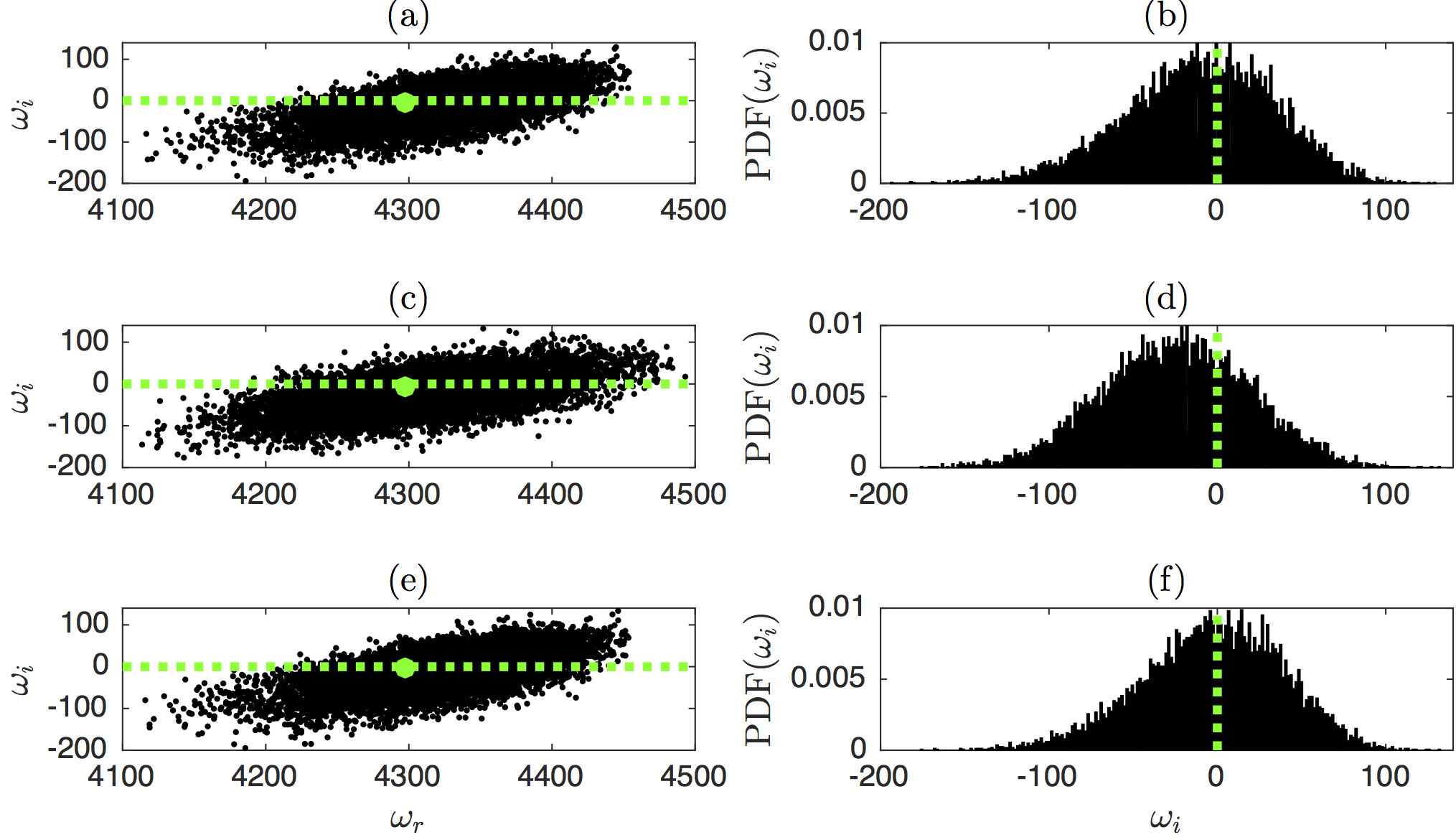}
  \caption{Same as Fig. \ref{fig:MCa} but for the strongly coupled Case C.  The big dot is the unperturbed deterministic eigenvalue, $\omega_0=4.304\times10^3rad/s+\mathrm{i}3.6\times10^{-3}s^{-1}$. The standard deviation of the uniform distribution of the flame parameters is $10\%$ (last row of Table \ref{tab:RF} for Case C).}\label{fig:MCc}
\end{figure}

The same quantities for Case C are shown in Fig. \ref{fig:MCc}. For this case, the risk factor predicted by MC is 42.2\%, by first-order AD is 30\%  and by second-order AD is $41.5\%$. 
 The strongly coupled configuration C is more prone to being unstable in practice than the weakly coupled configuration A:  For the same level of uncertainty of the flame parameters ($10\%$), the growth rate is uncertain up to within $\sim 150 s^{-1}$ in Case C, whereas it is uncertain up to within $\sim 20 s^{-1}$ in Case A.  

In general, the UQ analysis shows that the uncertainty present in the flame parameters can significantly affect the thermo-acoustic stability. Deterministic calculations of the eigenvalue (big circles in the left panels of Figs. \ref{fig:MCa},\ref{fig:MCc}) are not sufficient for a robust thermo-acoustic stability analysis, i.e., systems that are deterministically stable can have a great probability of becoming unstable. 
This is because thermo-acoustic systems are highly sensitive to changes in some design parameters, as shown in \cite{Magri2016jcp1}. Note that other sources of uncertainties, such as partly unknown acoustic losses, can affect the stability, although this is not considered here for simplicity. 

   \section{Uncertainty quantification via Monte Carlo method with Active Subspace Identification}\label{sec:asm}
Active Subspace Identification (ASI) is a method to reduce the parameter space dimension and create algebraic surrogate models useful to apply the Monte Carlo method for uncertainty quantification \cite{Constantine2014,Bauerheim2016}. 
The aim of this section is to combine the Monte Carlo method and ASI with the adjoint framework in order to further reduce the number of operations to perform. 
 
\subsection{The algorithm}\label{sec:asi_algo} 
\begin{sidewaysfigure}[!htb]
        \centering
        \includegraphics[width=1.0\linewidth]{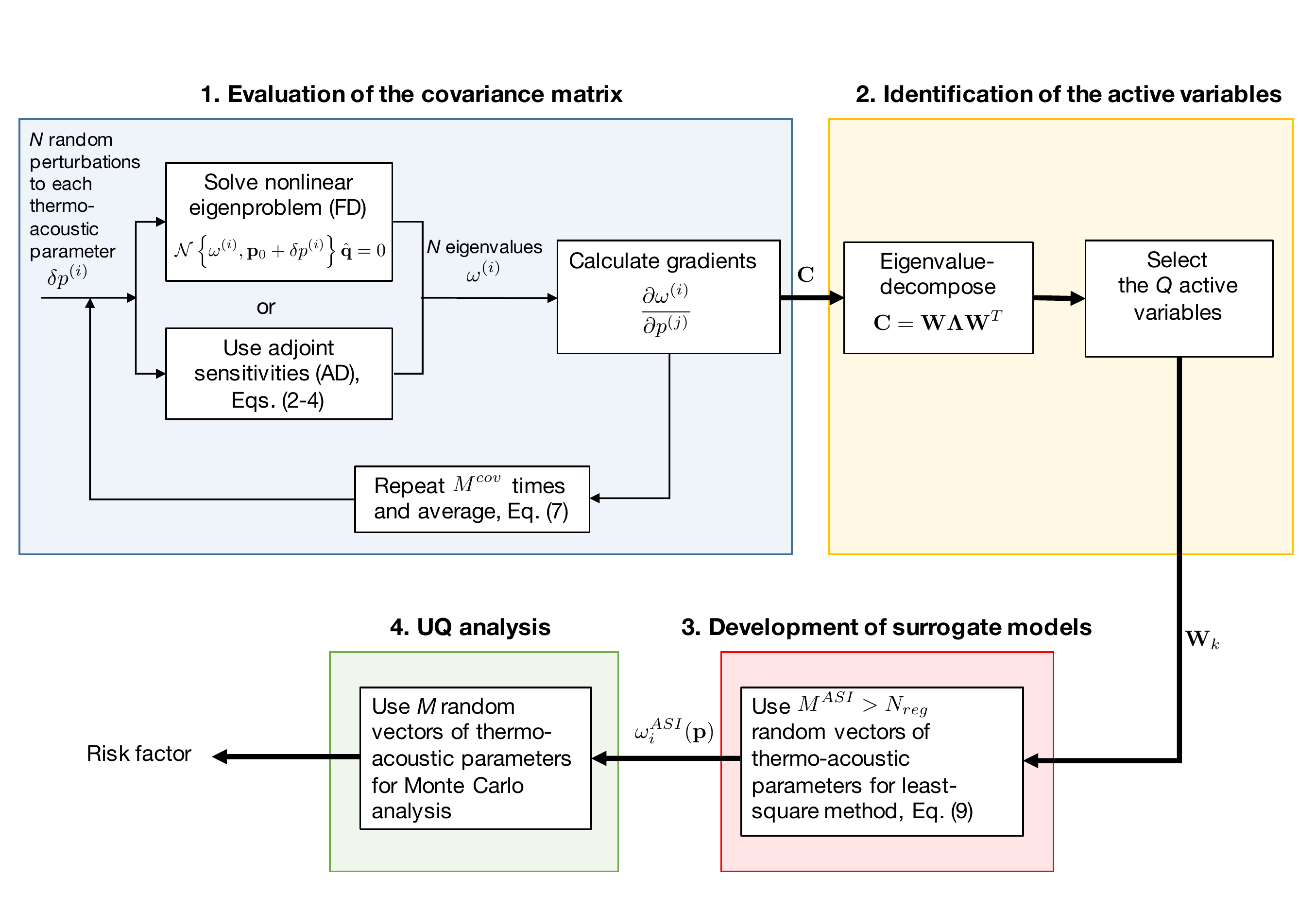}
  \caption{Active Subspace Identification combined with the Monte Carlo method for the calculation of the risk factor. The adjoint method (AD) avoids the calculation of the nonlinear eigenproblems required by the \lm{finite-difference method (FD)}.}\label{fig:asi_schem}
\end{sidewaysfigure}

Following \cite{Constantine2014,Bauerheim2016} and integrating the algorithm with an adjoint method, the procedure we propose to reduce the number of parameters by recognizing the active variables and applying the Monte Carlo method for UQ is as follows. 
\begin{enumerate}
\item \textbf{Evaluation of the covariance matrix}. 
We define the uncentred covariance matrix through the dyadic product 
\begin{align}\label{eq:acs1}
\mathbf{C} = \mathbb{E} [\mathbf{\nabla_{p}}\omega (\mathbf{\nabla_{p}} \omega)^T],
\end{align}
where the column vector $\mathbf{\nabla_{p}}\omega=[\partial\omega/\partial p_1 \; \partial\omega/\partial p_2 \; \ldots \; \partial\omega/\partial p_N]^T$ is the eigenvalue's sensitivity with respect to the $N$ thermo-acoustic parameters, and $\mathbb{E}$ is the expectation operator. Note that this vector consists of partial derivatives, therefore, $N$ eigenvalues need to be calculated.  
To compute the covariance matrix, we perform a Monte Carlo integration \cite{Constantine2014}, yielding
\begin{align}\label{eq:acs3}
\mathbf{C} \approx \frac{1}{M^{cov}}\sum_{j=1}^{M^{cov}} [\nabla_\mathbf{{p}}\omega(\mathbf{{p}}^{(j)}) (\nabla_\mathbf{{p}} \omega(\mathbf{{p}}^{(j)}))^T],
\end{align}
where the vector of parameters, $\mathbf{{p}}^{(j)}$, is drawn from the relevant uniform PDF of $M^{cov}$ Monte Carlo samples. As far as the number of computations is concerned, 
$M^{cov}\times N$ eigenvalues, $\omega$, are calculated by either \lm{finite difference (FD)}, which requires solving $M^{cov}\times N$ nonlinear eigenproblems, or the adjoint approach (AD), which requires solving only one nonlinear eigenproblem and its adjoint regardless of the number of parameters and Monte Carlo samples. 
\item \textbf{Identification of the active variables}. $\mathbf{C}$ is symmetric and, therefore, admits the real eigenvalue decomposition 
\begin{align}\label{eq:acs2}
\mathbf{C} =  \mathbf{W}\mathbf{\Lambda}\mathbf{W}^T.
\end{align}
Based on the relative importance of  the eigenvalues $\lambda_j$, we select the $Q$ dominant eigenvectors, $\mathbf{W}_k$. This choice might be rather subjective depending on the case \citep{Constantine2014}.  Fig.~\ref{fig:ASI_spec} shows that there are gaps between the first and second eigenvalues as well as  between the fifth and sixth eigenvalues. This suggests that five active variables should be kept. Physically, the first group ($\lambda_1$) is associated with a mean-flame effect, while the second group (from $\lambda_2$ to $\lambda_5$) corresponds to a symmetry-breaking splitting effect, as explained by \cite{Bauerheim2016}.

    \begin{figure}[!htb]
        \centering
        \includegraphics[width=0.9\linewidth]{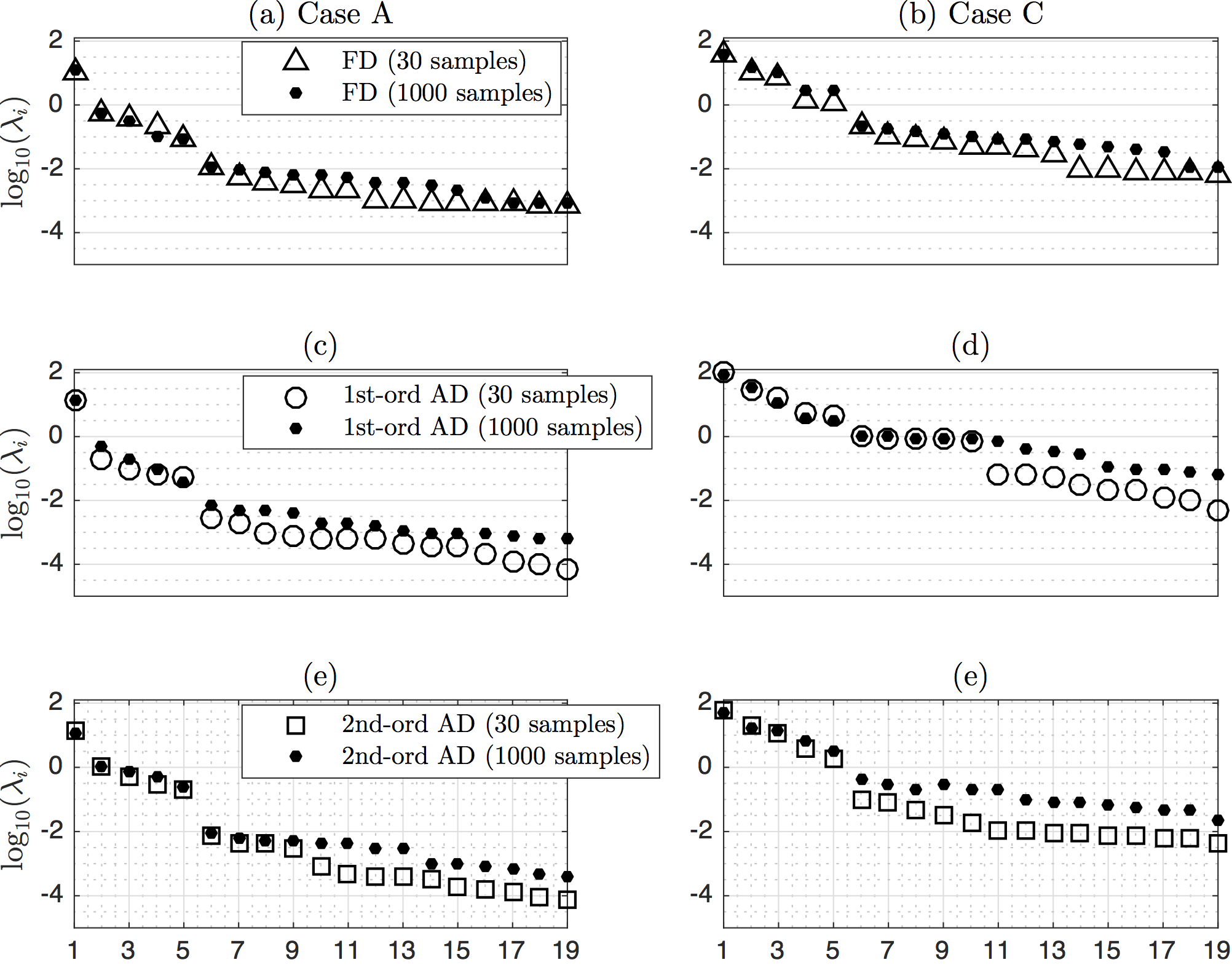}
  \caption{Part of the spectrum of the covariance matrix, $\mathbf{C}$. The dominant eigenvectors provide the $Q$ directions in the parameter space along which the growth rate varies the most. We recognize the first $Q=5$ eigenvalues as active variables because they are dominant and less sensitive to the Monte Carlo sampling, $M^{cov}$. Case A is shown in the left panels and Case C in the right panels.}\label{fig:ASI_spec}
\end{figure}

\item \textbf{Development of surrogate models.} We develop algebraic surrogate models for the growth rate as a function of the active variables, $\mathbf{W}^T_k\mathbf{p}$, $k=1,...,Q$, by a least-square method

\begin{align}
\omega_i^{ASI}(\mathbf{p})=\alpha_0&+\overbrace{\sum_{j=1}^{Q}\alpha_j\mathbf{W}^T_j\mathbf{p}}^{\mathrm{Linear}}+\ldots\nonumber\\
+&\underbrace{\sum_{j=1}^{Q}\sum_{k=j}^{Q}\alpha_{jk}(\mathbf{W}^T_j\mathbf{p})(\mathbf{W}^T_k\mathbf{p})
+\sum_{j=1}^{Q}\sum_{k=j}^{Q}\sum_{l=k}^{Q}\alpha_{jkl}(\mathbf{W}^T_j\mathbf{p})(\mathbf{W}^T_k\mathbf{p})(\mathbf{W}^T_l\mathbf{p})}_{\textrm{Quadratic and Cubic}}.  \label{eq:surr1}
\end{align}
The function $\omega_i(\mathbf{p})^{ASI}$ is also known as the response surface and needs $M^{ASI}$ Monte Carlo samples for least-square fitting, where $M^{ASI}$ is greater than the number of regression coefficients, $N_{reg}$. 
Here, we compare a linear regression model with a cubic one by (i) retaining only the active variables, $Q=5$ (Fig.~\ref{fig:ASI_spec}), and (ii) using all 38 variables, $Q=38$. A detailed comparison of different regression models is beyond the scope of this paper because the focus is on the adjoint methods. Other surrogate models were tested by \cite{Bauerheim2016}.

\item \textbf{UQ analysis}. $M$ cheap Monte Carlo algebraic evaluations of $\omega^{ASI}_i$ are performed to estimate the growth-rate risk factor, avoiding the nonlinear eigenproblem. In both Cases, we use  $M=50,000$, which ensures convergence of the risk factor and PDFs (not shown). 
\end{enumerate} 
This procedure is summarized in Fig.~\ref{fig:asi_schem}. 
\clearpage
\clearpage

\subsection{Results}\label{sec:asi_results}
\begin{figure}[!htb]
        \centering
        \includegraphics[width=0.9\linewidth]{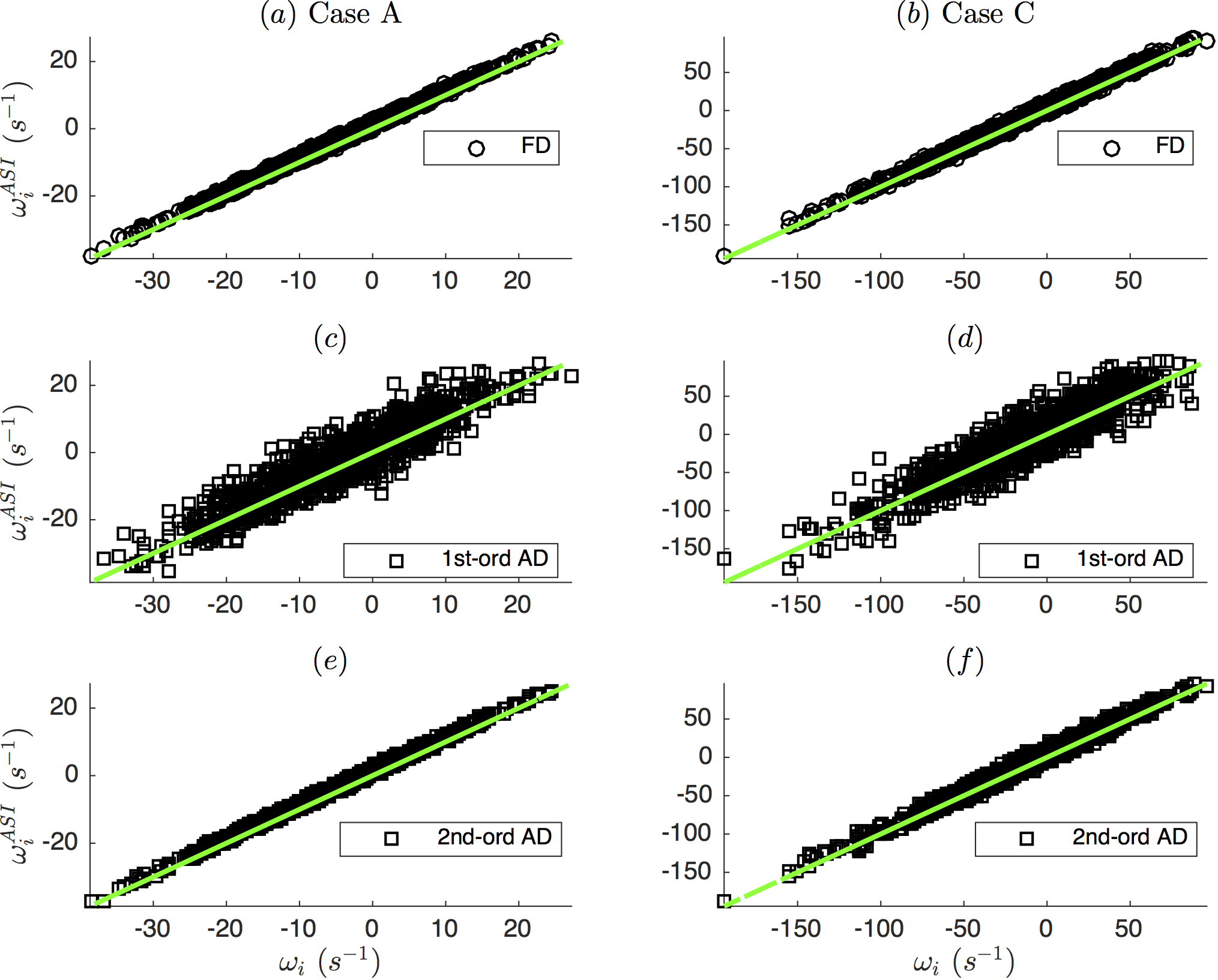}
  \caption{50,000 Monte Carlo experiments  obtained by evaluating the cubic least-square surrogate models by ASI with $Q=5$ active variables.  Results from  \lm{FD}-based (first row);  1st-order AD-based (second row) and 2nd-order AD-based (third row) surrogate models. Case A is shown on the left and Case C on the right. The straight line is the correct solution, the higher the scattering, the larger the error.}\label{fig:ASI_AC}
\end{figure}

First, we investigate the accuracy of the surrogate models obtained by \lm{FD} and AD methods. 
In Fig. \ref{fig:ASI_AC}, the results are shown for the weakly coupled Case A and strongly coupled Case C. In these charts, the straight line represents the correct perturbed growth rate: the larger the scattering of the growth rates calculated via ASI surrogate models, the larger the error. The scattering is quantified by the coefficients of determination, reported  in Table \ref{tab:ASI}, defined as 
\begin{align}
& R^2 =1-\frac{\sum_{i=1}^{M}(\omega^{ASI}_i - \bar{\omega}_i)^2}{\sum_{i=1}^{M}(\omega_i-{\bar{\omega}}_i)^2},
\end{align}
where the bar  indicates the mean value of the MC database of Section~\ref{sec:monca} ($M=10,000$) and the superscript $ASI$ indicates the eigenvalue obtained by running the Monte Carlo UQ analysis with the surrogate models.  
The scattering between the \lm{FD}-surrogate models and the 2nd-order AD-surrogate models (equation~\eqref{2eq:eigpert8}) is similar (Table~\ref{tab:ASI}), meaning that the AD method can be applied to ASI. The accuracy obtained via ASI does not increase significantly as the number of eigenvectors is retained, which means that the variables retained are indeed the most influential (active). For the configurations analysed, the 1st-order AD-surrogate model is less accurate. 

Secondly, the results of the uncertainty quantification are reported in Table \ref{tab:ASI}, which shows the risks factors calculated with  the \lm{FD}-based models and  the AD-based models. The correct growth-rate risk factor calculation is given by MC from the standard Monte Carlo simulation (first row of Table \ref{tab:ASI}). 
As for the number of computations, applying ASI with \lm{finite difference} involves calculating $M^{cov}\times N$$=$$30\times 38$$=$$1,140$ eigenproblems. On the other hand, by using the adjoint approach we have to solve only one nonlinear eigenproblem and its adjoint, and use the sensitivity equations of Section \ref{sec:appannu}. 
The second-to-last column of Table \ref{tab:ASI} summarizes the number of nonlinear-eigenproblem computations needed following \lm{FD} and AD surrogate models and the MC method of Section \ref{sec:monca}. 

Thirdly, we discuss the complexity and accuracy of the surrogate models. These algebraic models are computationally cheap providing that the number of regression coefficients is small. This number is exactly given by
\begin{align}
N_{reg} = 1 +Q +\phi\left[ 2Q + \frac{3}{2}Q(Q-1)+\frac{1}{6}Q(Q-1)(Q-2) \right], 
\end{align}
where $\phi=0$ in linear regression and $\phi=1$ in cubic regression. The last column of Table \ref{tab:ASI} reports $N_{reg}$ for the four surrogate models developed. 
On the one hand, linear regression is cheap because it requires only $N_{reg}=6$ but the predicted risk factors and coefficients of determination are unsatisfactory. This does not appreciably improve when all the $Q=38$ ($N_{reg}=39$) variables are retained. 
On the other hand, a cubic regression is needed to obtain good predictions of the risk factor. Retaining all the variables, $Q=38$, needs $N_{reg}=10,660$, which makes the development of such a surrogate model computationally time-consuming. However, retaining only the $Q=5$ active variables requires the calculation of only $N_{reg}=56$, which significantly decreases the complexity of the original Monte Carlo problem keeping high accuracy on the risk factors. In summary, the cubic regression model\footnote{Quadratic models are not considered here because Bauerheim et al. \cite{Bauerheim2016} showed that modelling the cubic terms appreciably improves the accuracy of the response surface.} based on the first five active variables provides an excellent compromise between computational cost and accuracy, also when the gradients are obtained by the adjoint method.  

\clearpage
\begin{table}[!htb]
\centering
\renewcommand{\arraystretch}{1.3} 
 \small{ \begin{tabular}{ c | c |l | c |  c | c | c | c | c}
 \multicolumn{1}{c }{} &\multicolumn{1}{c }{} & & \multicolumn{2}{c |}{Risk factor} & \multicolumn{2}{|c|}{$R^2$}& & \\ \cline{4-7}
   \multicolumn{1}{c }{}  &    &Model & Case A  & Case C & Case A & Case C & \makecell{Nonlinear\\ eigenproblems} & $N_{reg}$\\ \Cline{2pt}{3-9}
   \multicolumn{1}{c }{} &            & MC (Tab. \ref{tab:RF})& 34.5\% & 42.2\% & 1& 1& $M=10,000$ & --\\\Cline{2pt}{2-9}

 \parbox[t]{2mm}{\multirow{6}{*}{\rotatebox[origin=c]{90}{Linear regression}}}   & \parbox[t]{2mm}{\multirow{3}{*}{\rotatebox[origin=c]{90}{$Q=38$}}}        & \lm{FD}                     &    38\%                          &  44.4\%   & 0.81 & 0.83 & $M^{cov}\times N = 1,140$  &\\  
  &      &1st-order AD      &     45\%                          & 49.8\% & 0.72&0.75 & 1  & 39\\ 
   &     &2nd-order AD    & 39.4\% & 45.3\% &  0.80& 0.81& 1 &\\  \Cline{1pt}{2-9}
 & \parbox[t]{2mm}{\multirow{3}{*}{\rotatebox[origin=c]{90}{$Q=5$}}}          &\lm{FD}                       &       42\%                      &  45.8\% &  0.80& 0.80& $M^{cov}\times N = 1,140$  & \\  
 &       & 1st-order AD       &      53.2\%                  & 50.3\% & 0.7&0.77 & 1  & 6\\  
  &      &2nd-order AD      & 40.8\% & 45.1\%& 0.79& 0.83&1 & \\ \Cline{2pt}{1-9}
  
 \parbox[t]{2mm}{\multirow{6}{*}{\rotatebox[origin=c]{90}{Cubic regression}}}   & \parbox[t]{2mm}{\multirow{3}{*}{\rotatebox[origin=c]{90}{$Q=38$}}}        & \lm{FD}                     &    35\%                          &  42.9\%   & \textbf{0.96} & \textbf{0.94} & $M^{cov}\times N = 1,140$  &\\  
  &      &1st-order AD      &     40.2\%                          & 47.3\% & 0.86&0.82 & 1 &  10,660\\ 
   &     &2nd-order AD    & 35.6\% & 43.6\% &  \textbf{0.95}& \textbf{0.94} & 1 &\\  \Cline{1pt}{2-9}
 & \parbox[t]{2mm}{\multirow{3}{*}{\rotatebox[origin=c]{90}{$Q=5$}}}          
& \lm{FD}                     &    35.1\%                          &  43.1\%   & \textbf{0.95} & \textbf{0.94} & $M^{cov}\times N = 1,140$ &  \\  
  &      &1st-order AD      &     44.5\%                          & 46.2\% & 0.85&0.83 & 1  & 56\\ 
   &     &2nd-order AD    & 35\% & 43.3\% &  \textbf{0.95}& \textbf{0.94} & 1 &\\  
  \end{tabular}}
\caption{Growth-rate risk factors and number of nonlinear eigenproblems by \lm{finite-difference (FD)} and adjoint- (AD) based surrogate models for the weakly coupled and strongly coupled Cases A and C. Linear and cubic least-square surrogate models based on $Q=5$ and $Q=38$ active variables. The coefficient of determination, $R^2$, is a measure of the scattering of Fig. \ref{fig:ASI_AC}, hence, the accuracy of the surrogate model ($R^2>90\%$ are highlighted in bold). The results for the cubic surrogate model with $Q=5$ are shown in Fig.~\ref{fig:ASI_AC}.}
\end{table}\label{tab:ASI}
\clearpage

\begin{figure}[!htb]
        \centering
        \includegraphics[width=0.9\linewidth]{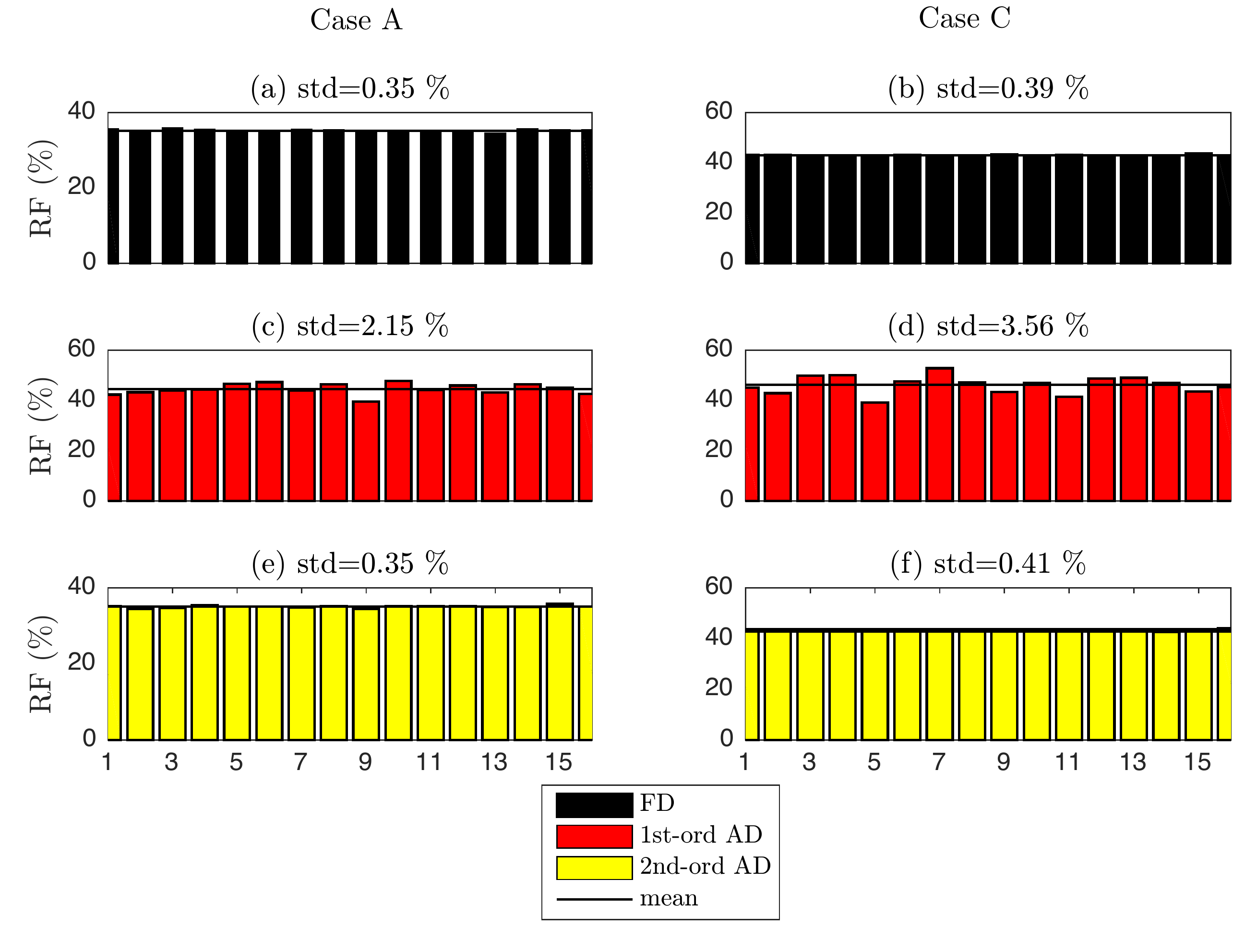}
  \caption{Risk factors obtained by running 16 sets of $M=50,000$ Monte Carlo simulations with the cubic surrogate model with $Q=5$. Case A is depicted in the left panels, Case C in the right panels. `std' stands for standard deviation. }\label{fig:var_AC}
\end{figure}

Finally, to evaluate how robust the calculation of the risk factor is, we run 16 sets of $M=50,000$ Monte Carlo simulations and calculate the mean and standard deviations of the results, as shown in Fig.~\ref{fig:var_AC} for the cubic surrogate model with $Q=5$. 
Repeating this procedure several times provides an estimation of the confidence interval of the risk factor. 
As shown in Table \ref{tab:ASI}, the first-adjoint method is unsatisfactorily accurate in the weakly coupled regime (RF $=$ $44.5\%$ compared with RF $=$ $35.1\%$ by MC) but in the strongly coupled regime the estimation is more accurate (RF $=$ $46.2\%$ compared with RF $=$ $43.1\%$ by MC). However, Fig.~\ref{fig:var_AC} reveals that the latter estimation is not robust because the interval of confidence by the first-order adjoint method is $\pm 3.56\%$ (panel d). Nevertheless, using the second-order adjoint method (panel f) provides a reliable risk factor because the mean value RF=$43.3\%$ is in agreement with \lm{FD} (panel b) and the standard deviation is small, $\pm 0.41\%$. 
This analysis indicates that combining a second-order adjoint method with the ASI method to compute surrogate models is a robust and accurate method to predict stability margins of annular combustors. 

%
\section{Conclusions} 
Deterministic calculations of the growth rates of two annular-combustor configurations are not sufficient for a robust thermo-acoustic stability analysis. It is shown that systems that are deterministically stable can have a great probability of becoming unstable because thermo-acoustic systems are highly sensitive to changes in some design parameters. 
In order to calculate the probabilities that the annular combustors are unstable (risk factors), given uncertainties in the flame parameters, we first combine an adjoint algorithm  with a standard Monte Carlo method. The risk factors and probability density functions are accurately predicted by adjoint methods and the number of nonlinear eigenproblems solved is reduced by a factor equal to the number of Monte Carlo samples which, in this case, is 10,000. The strongly coupled annular combustor (Case C), of industrial interest, is found to be significantly more sensitive to uncertainties of the flame parameters than the weakly coupled configuration (Case A). 

Secondly, we combine the adjoint algorithm with Active Subspace Identification to develop growth-rate algebraic models to further reduce the number of computations required by the standard Monte Carlo method. The  number of nonlinear eigenproblems solved is reduced by a factor equal to the number of Monte Carlo samples needed to calculate the covariance matrix, which, in this study is 1,140. The surrogate model obtained by cubic regression is found to be an excellent compromise between computational cost and numerical accuracy.

The first-order adjoint framework is a reliable tool for uncertainty quantification as long as the rate of change of eigenvalues with parameters is approximately linear around the operating point in question and, when it is not,  the standard deviations of the system's parameters are not too large. 
In these scenarios, the second-order adjoint method proved more accurate and versatile. 
The adjoint framework is a promising method for design to obtain quick accurate estimates of risk factors at very cheap computational cost.

\subsection*{Acknowledgments}

The authors are grateful to the 2014 Stanford Center for Turbulence Research Summer Program (Stanford University) where the ideas of this work were born. 
L.M. and M.P.J acknowledge the European Research Council - Project ALORS 2590620 for financial support. Discussions with Wolfgang Polifke and Camilo Silva are gratefully appreciated. 
\section*{References}
                 \bibliographystyle{elsarticle-num} 
                 \bibliography{MyLibrary}

\end{document}